\documentclass[conference]{IEEEtran}
\IEEEoverridecommandlockouts
\usepackage{cite}
\usepackage{amsmath,amssymb,amsfonts}
\usepackage{algorithmic}
\usepackage{graphicx}
\usepackage{textcomp}
\usepackage{xcolor}
\usepackage{orcidlink}
\usepackage{makeidx}
\usepackage{marvosym} % for \Letter symbol
\usepackage{graphicx}
\usepackage{amsmath}
\usepackage{multirow}
\usepackage{hyperref}
\definecolor{darkgreen}{rgb}{0.0, 0.5, 0.0}
\def\BibTeX{{\rm B\kern-.05em{\sc i\kern-.025em b}\kern-.08em
    T\kern-.1667em\lower.7ex\hbox{E}\kern-.125emX}}
    
\begin{document}

\title{Hard-Attention Gates with Gradient Routing for Endoscopic Image Computing}

\author{\IEEEauthorblockN{Giorgio Roffo ~\Letter~\orcidlink{0000-0003-4170-914X}}
\IEEEauthorblockA{\textit{Cosmo IMD}\\
Dublin, Ireland \\
\textit{groffo@cosmoimd.com}
}
\and
\IEEEauthorblockN{Carlo Biffi~\orcidlink{0000-0002-4913-7441}}
\IEEEauthorblockA{\textit{Cosmo IMD}\\
Dublin, Ireland \\
cbiffi@cosmoimd.com}
\and
\IEEEauthorblockN{Pietro Salvagnini~\orcidlink{0000-0002-1103-894X}}
\IEEEauthorblockA{\textit{Cosmo IMD}\\
Dublin, Ireland \\
psalvagnini@cosmoimd.com}
\and
\IEEEauthorblockN{Andrea Cherubini ~\Letter~\orcidlink{0000-0002-5946-4390}}
\IEEEauthorblockA{\textit{Cosmo IMD}\\
Dublin, Ireland \\
acherubini@cosmoimd.com}
}

\maketitle

\begin{abstract}
To address overfitting and enhance model generalization in gastroenterological polyp size assessment, our study introduces \textit{Feature-Selection Gates (FSG)} or \textit{Hard-Attention Gates (HAG)} alongside Gradient Routing (GR) for dynamic feature selection. This technique aims to boost Convolutional Neural Networks (CNNs) and Vision Transformers (ViTs) by promoting sparse connectivity, thereby reducing overfitting and enhancing generalization. HAG achieves this through sparsification with learnable weights, serving as a regularization strategy. GR further refines this process by optimizing HAG parameters via dual forward passes, independently from the main model, to improve feature re-weighting. Our evaluation spanned multiple datasets, including CIFAR-100 for a broad impact assessment and specialized endoscopic datasets (REAL-Colon~\cite{realcolon2023}, Misawa~\cite{misawa2021cad}, and SUN~\cite{suncolon2020}) focusing on polyp size estimation, covering over 200 polyps in more than 370K frames. The findings indicate that our HAG-enhanced networks substantially enhance performance in both binary and triclass classification tasks related to polyp sizing. Specifically, CNNs experienced an F1 Score improvement to 87.8\% in binary classification, while in triclass classification, the ViT-T model reached an F1 Score of 76.5\%, outperforming traditional CNNs and ViT-T models. 
To facilitate further research, we are releasing our codebase, which includes implementations for CNNs, multistream CNNs, ViT, and HAG-augmented variants. This resource aims to standardize the use of endoscopic datasets, providing public training-validation-testing splits for reliable and comparable research in gastroenterological polyp size estimation. The codebase is available at \href{http://github.com/cosmoimd/feature-selection-gates}{github.com/cosmoimd/feature-selection-gates}.
\end{abstract}


\section{Introduction}
\begin{figure*}[t]
    \centering
    \includegraphics[width=1.0\linewidth]{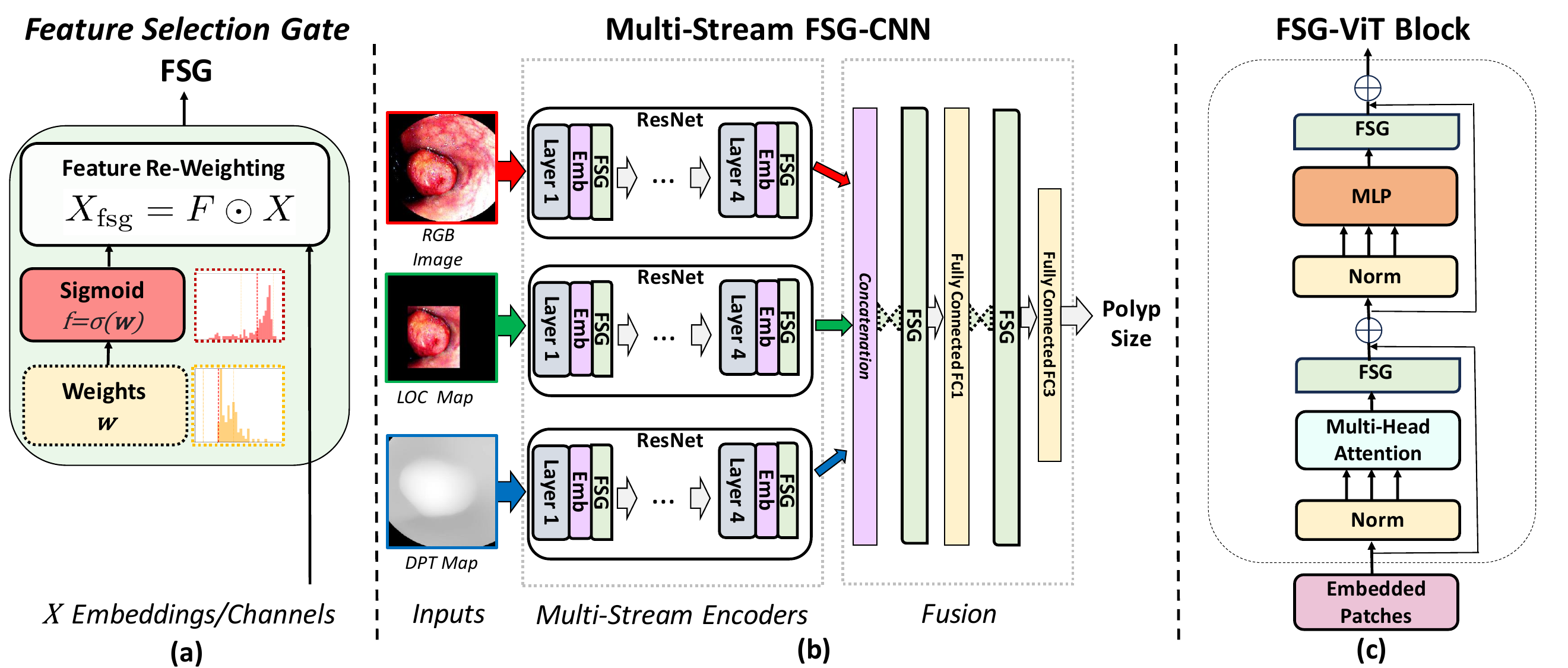}
    \caption{Feature Selection-Attention Gates (HAG) Integration in Deep Learning Models. (a) Conceptual design of HAG. (b) Application of HAG in a multi-stream CNN architecture, with each stream being optional. (c) Embedding of HAG within a ViT block, positioned after the multihead attention and the MLP for enhanced feature re-weighting.}
    \label{fig:fsgnet}
\end{figure*}
Deep learning (DL) techniques, particularly convolutional neural networks and transformers, have significantly advanced the analysis of endoscopic imaging~\cite{miccai4} \cite{miccai1} \cite{miccai2} \cite{miccai3}. Nonetheless, these models encounter challenges such as overfitting when applied to the typically smaller datasets found in endoscopy, in contrast to larger datasets like ImageNet. These challenges stem from various factors, including privacy concerns, the need for expert annotations, associated costs, and the inherent variability of endoscopy imaging modalities. Moreover, the infrequency of certain conditions, such as large colorectal polyps, intensifies data imbalance issues, further complicating the development of reliable and precise DL models for endoscopic image analysis~\cite{itoh2021binary} \cite{itoh2018towards} \cite{sudarevic2023artificial} \cite{abdelrahim2022automated} \cite{atalaia2019variation} \cite{duffield2005learn}.

A review of related work on polyp size estimation underscores the critical nature of accurate assessment for the effective management and surveillance of colorectal cancer~\cite{PopescuCrainic2024}. The size of a polyp significantly influences its potential for malignancy, necessitating accurate measurement. According to guidelines from prominent endoscopy organizations, including the American and European Societies for Gastrointestinal Endoscopy (\textit{ASGE} and \textit{ESGE}), polyps can be classified in D-S-L categories based on their size: (D) Diminutive polyps (5mm or smaller), (S) Small polyps (between 5mm and 10mm), and (L) Large polyps (10mm or larger). The societies particularly emphasize the removal and subsequent histopathological examination of Large polyps due to their increased risk of cancer, underscoring the critical role of precise size determination~\cite{Hassan2013,Ferlitsch2017}. However, there is notable variability in manual size estimations by endoscopists~\cite{atalaia2019variation}, which poses a risk of mismanagement~\cite{Gupta2011}. The development of automated estimation techniques via computer vision aims to mitigate this issue, providing more consistent measurements through machine learning models trained on a diverse array of imaging data~\cite{PopescuCrainic2024}. In pursuit of standardizing polyp size estimation, our work leverages state-of-the-art 2D and 3D estimation methodologies~\cite{abdelrahim2022automated,sudarevic2023artificial,itoh2022uncertainty,batlle2023lightneus} and enhances dataset diversity~\cite{realcolon2023,misawa2021cad,suncolon2020}. This dual approach aims to refine the accuracy of automated polyp-size estimation, addressing both the technological and data-related challenges inherent in this domain.

Building upon this foundation, we introduce Feature Selection-Attention Gates (HAG) and Gradient Routing (GR). These innovative mechanisms, tailored for the gastroenterological polyp size assessment domain, counter overfitting and enhance model generalization. Drawing from ~\cite{yu2018_NISP,Roffo:InfFS:2015,roffo2016feature,roffo2016online,verikas2002feature}, our approach promotes sparse connectivity in deep networks and uses a dual forward pass strategy for gradient routing. This fosters model sparsity and efficiency while selectively emphasizing pertinent features~\cite{yu2018_NISP,roffo2020infinite}. Figure \ref{fig:fsgnet} illustrates the integration of HAG in DL frameworks. Fig. \ref{fig:fsgnet}(a) outlines the HAG's conceptual design, employing sigmoid-normalized weights ranging between 0 and 1. Fig. \ref{fig:fsgnet}(b) integrates HAG into CNNs in a multi-stream setup, accommodating various input types like RGB, Depth, and Location Maps. Fig. \ref{fig:fsgnet}(c) illustrates HAG's incorporation in the ViT model, placing one HAG after the multihead attention and another following the MLP block. 

The proposed approach was evaluated, with our primary focus on public endoscopic datasets, consisting of 232 polyps across more than 370K frames in the REAL-Colon~\cite{realcolon2023}, Misawa~\cite{misawa2021cad} and the SUN database \cite{suncolon2020}. These databases encompass realistic clinical scenarios, such as variable lighting and obstructions. Additionally, to assess the impact of HAG on ViT performance in a more general context, we also conducted evaluations on CIFAR-100. The HAG ResNet18 (R18) and ViT models showed improved accuracy, achieving an accuracy of 75.2\% and 83.8\% respectively, outperforming by +1.3\% and +5.8\% their baselines and SotA~\cite{TinyViT2022,Deng2020R18Cifar}. 
%In the experiments on polyp size estimation we compare the methods with the SotA including a range of inputs like RGB, Depth~\cite{MIDAS}, and Location Maps, implemented in multi-stream architectures. 
In polyp size estimation experiments, we compared methods with SotA using RGB, Depth~\cite{MIDAS}, and Location Maps in CNNs with Dropout and Batch Normalization. Hybrid methods like CBAM \cite{woo2018cbam} were excluded to focus on pure CNN and transformer architectures.

In the bias-variance tradeoff analysis, HAG models showed superior performance with higher F1 scores and average sensitivity-specificity in a 6-fold experiment. 
In a separate, consolidated evaluation across all dataset folds, HAG models consistently outperformed standard models in both binary~\cite{itoh2021binary,itoh2018towards} (under and over 10 mm polyps) and triclass classifications (Diminutive-Small-Large). 

Across 12 methods compared, the highest average performances were from \textbf{HAG MultiStream-R18 (LOC+DPT)} at \textbf{66.1\%} and \textbf{HAG ViT-T (RGB)} at \textbf{65.5\%}. Our findings indicate that ViT models without LOC maps or DPT are preferred due to reduced error propagation. ViT Tiny, with 5.6M parameters and 4.7G flops, is the most efficient. HAG integration significantly enhances ViT's performance in regression and classification. Our unique integration of the following components to address overfitting in medical image analysis is a key innovation. The main contributions of our work can be summarized as follows:
\begin{enumerate}
\item \textbf{Hard-Attention Gate (HAG)}: Acts as an online regularization tool to enhance learning, reduce overfitting, and improve generalization.
\item \textbf{Gradient Routing (GR)}: Optimizes HAG parameters separately from the main model, allowing tailored learning rates and gradient clippings.
\item \textbf{Enhancing Vision Transformers and CNNs}: HAG enhances ViTs and CNNs, including multi-input variants, for versatile processing of RGB, Depth maps~\cite{MIDAS}, and location maps.
% \item \textbf{Adapted CNNs for Multiple Inputs}: Processes various inputs like RGB, Depth by ~\cite{MIDAS}, and Location Maps effectively.
% \item \textbf{Dataset Standardization}: Includes uniform dataset splits and accessible code for reproducible and comparable research.
% \item \textbf{Performance Benchmarking}: Establishes new field benchmarks and promotes fair comparisons through official splits and shared HAGNet code.
\end{enumerate}
% Our contribution integrates public datasets and provides code and standardized splits, ensuring reproducibility and consistency in polyp size estimation research.

\section{Methodology}

% In this section, we introduce the HAG module (see Fig.\ref{fig:fsgnet}) and Gradient Routing strategy for FSG training.

\subsection{Feature Selection-Attention Gates (HAG)}

The HAG (in Fig.\ref{fig:fsgnet}) dynamically assigns weights to each feature within the model, applicable to embeddings and channels in architectures such as Transformers and CNNs. These weights, dynamically adjustable during the training process, are normalized using a sigmoid function to ensure values range between 0 and 1. This weighting mechanism facilitates focused learning, allowing the model to prioritize more informative features and reduce the less relevant ones.

% For the set of input features \( X = (x_1, x_2, \ldots, x_n) \), the feature selection weights, denoted as \( F = (f_1, f_2, \ldots, f_n) \), are obtained by applying a $sigmoid$ function over the raw feature weights \( W = (w_1, w_2, \ldots, w_n) \) obtaining the FSG-scores. These FSG scores are then applied to the embedding via the Hadamard product. The process can be encapsulated in the following equation:
% \begin{equation}
% X_{\text{fsg}} = F \odot X = (f_1=\sigma(w_1) \cdot x_1, f_2=\sigma(w_2) \cdot x_2, \ldots, f_n=\sigma(w_n) \cdot x_n)
% \end{equation}
Given a set of input features \(X\), represented as \(X = (x_1, x_2, \ldots, x_n)\), where \(n\) is the number of features or the embedding size, and each \(x_i\) for \(i=1, 2, \ldots, n\) corresponds to a specific feature in the input data. Feature selection is performed by introducing a set of weights \(F\), denoted as \(F = (f_1, f_2, \ldots, f_n)\). These weights are derived from the raw feature weights \(W = (w_1, w_2, \ldots, w_n)\) through a transformation. Specifically, we apply a sigmoid function \(\sigma\) to each raw weight \(w_i\) to obtain the corresponding FS weight \(f_i\). The sigmoid function, defined as \(\sigma(z) = \frac{1}{1 + e^{-z}}\) for any input \(z\) makes the weights suitable for FS by scaling them between 0 and 1. These transformed weights are referred to as HAG-scores.

The HAG-scores are then applied to the original input features \(X\) to obtain the relevant version of the input, denoted as \(X_{\text{att}}\). This application is performed through the Hadamard product of \(F\) and \(X\), effectively scaling each feature \(x_i\) by its corresponding HAG-score \(f_i\) (see Fig.\ref{fig:fsgnet}). Mathematically, this process is encapsulated in the following equation:
\begin{equation}
\begin{split}
X_{\text{att}} &= F \odot X = f_1 \cdot x_1, f_2 \cdot x_2, \ldots, f_n \cdot x_n\\
&= \sigma(w_1) \cdot x_1, \sigma(w_2) \cdot x_2, \ldots, \sigma(w_n) \cdot x_n,
\end{split}
\end{equation}

where \(\odot\) denotes the Hadamard product, and each \(f_i=\sigma(w_i)\) is the result of applying the sigmoid function to the raw weight \(w_i\), which is then multiplied by the corresponding input feature \(x_i\) to achieve the feature selection effect.

Unlike attention mechanisms that use softmax for input weighting, HAG \textit{independently} re-weights features with scores from 0 to 1, not summing to 1. This enables synergy with attention in ViT architectures, enhancing performance (see HAG-GR weight distributions in supplementary material).

\subsection{Gradient Routing for Online Feature Selection}

Gradient Routing (GR) in our model employs a dual-phase optimization approach with distinct optimizers for different model components. Initially, GR updates the HAG parameters. This step focuses on refining feature weights only. Following this, GR updates the main model parameters using a different optimizer, based on the adjusted state from the HAG phase. 
The iterative nature of GR aligns with the principles of gradient descent and backpropagation, starting with the fine-tuning of HAG parameters and then progressing to the main model parameters. The small derivatives introduced by the sigmoid function in deep layers can lead to vanishing gradients, minimally updating early layer weights. To counter this, a gradient clipping strategy with different thresholds for HAG and the main model can be used. Higher thresholds for HAG address the sigmoid's limitations, and lower ones for the main model ensure stability. This method ensures efficient backpropagation across the network~\cite{clipping2013}, optimizing learning in both HAG and the main model components. The GR method utilizes a dual-phase optimization with gradient clipping for HAG and main model parameters, diverging from the layer-wise pre-training and fine-tuning strategy described in ~\cite{Bengio2006}. The gradient updating process in GR can be represented as:
\begin{equation}
\theta_{\text{\textit{att}}}^{t+1} = \theta_{\text{\textit{att}}}^{t} - \eta_{\text{att}} \text{clip}(\nabla_\text{\textit{att}} L(\theta_{\text{\textit{att}}}^{t}, \theta_{\text{\textit{main}}}^{t},  D ), \text{Th}_{\text{\textit{att}}})
\end{equation}
\begin{equation}
\theta_{\text{\textit{main}}}^{t+1} = \theta_{\text{\textit{main}}}^{t} - \eta_{\text{\textit{main}}} \text{clip}(\nabla_{\text{\textit{main}}} L(\theta_{\text{\textit{main}}}^{t}, \theta_{\text{\textit{att}}}^{t+1}, D), \text{Th}_{\text{\textit{main}}})
\end{equation}
where \( \theta_{\text{main}}^{(t)} \) and \( \theta_{\text{att}}^{(t)} \) are the parameters of the main model and HAG at iteration \( t \), \( \eta_{\text{main}} \) and \( \eta_{\text{att}} \) are the respective learning rates, \( \nabla L \) denotes the gradient of the loss function, \( D \) is the training data, and \(\text{clip}(\cdot, \text{Th})\) is the gradient clipping function with specified thresholds.

\section{Experiments and Results}\label{sec:exps}

Images were resized to \(384 \times 384\) and normalized using dataset-specific mean and standard deviation computed on the training data, ensuring dataset-specific color adjustments. Circular cropping was used to isolate the central part of an image into a circular shape, thereby concentrating analysis on relevant areas and eliminating peripheral distractions. Standard data augmentation included rotations, color adjustments, and noise addition. Polyp sizes normalized to \([-1, +1]\) range were used for stable regression training. A domain-specific weighted Huber Loss addressed the imbalanced distribution of polyp sizes within the dataset:
\begin{equation}
\begin{minipage}[b]{0.5\linewidth}
    \centering
    $\textbf{A} = \begin{cases} 
    \alpha_1, & \text{if } T_1 < y \leq T_2 \\
    \alpha_2, & \text{if } y > T_2 \\
    1, & \text{otherwise}
    \end{cases}$
    \label{eq:A}
\end{minipage}%
\begin{minipage}[b]{0.5\linewidth}
    \centering
    $L_W = \frac{1}{N} \sum_{i=1}^{N} (\text{Huber}(x_i, y_i) \cdot \textbf{A}_i)$
\end{minipage}
\end{equation}
with $(T_1, \alpha_1) = (5, 2)$ and $(T_2, \alpha_2) = (10, 3)$, and $N$ representing the mini-batch size. The Adam optimizer was utilized, with a learning rate set in the range of \([10^{-3}, 10^{-5}]\), and weight decay specified within the interval \([10^{-5}, 10^{-8}]\). Gradient clipping was set between 5 to 10 for CNNs, and a cosine annealing scheduler with warm restarts was applied for learning rate control. Parameters within the HAG modules were initialized using the Xavier method. For ViT models, gradient clipping threshold was set to 128 to mitigate the vanishing gradient issue~\cite{clipping2013} (see Supplementary Material for experimental setup). 
We reported model parameters in Table~\ref{tab:bias_variance_analysis} and conducted experiments using a Tesla V100-PCIE GPU with 32GB memory. ResNet-18 and ViT-Tiny models operate in real-time, requiring approximately 2GB for training and inference with a batch size of 1 on 384x384x3 images. The HAG adds minimal parameters and an imperceptible increase in FLOPS, as detailed in Table~\ref{tab:bias_variance_analysis}.
\begin{table*}[t]
\centering
\caption{Distribution of Frames (\#Polyps) in Endoscopic Dataset Folds from REAL-Colon \cite{realcolon2023}, Misawa et al.'s Database \cite{misawa2021cad}, and the SUN Dataset \cite{suncolon2020}}
\label{tab:dataset_stats}
\resizebox{0.85\textwidth}{!}{%
\begin{tabular}{|c|c|c|c|c|c|c|}
\hline
\textbf{Categories} & \textbf{Fold 1} & \textbf{Fold 2} & \textbf{Fold 3} & \textbf{Fold 4} & \textbf{Fold 5} & \textbf{Fold 6} \\ \hline
\textit{Diminutive} & 46,982 (29) & 49,648 (30) & 44,640 (30) & 54,707 (30) & 49,256 (29) & 51,784 (29) \\ \hline
\textit{Small} & 9,535 (5) & 11,985 (6) & 6,098 (6) & 827 (5) & 1,255 (6) & 4,538 (4) \\ \hline
\textit{Large} & 6,298 (5) & 24,189 (3) & 1,086 (3) & 6,459 (4) & 1,105 (3) & 1,648 (5) \\ \hline
\textbf{Total} & 62,815 (39) & 85,822 (39) & 51,824 (39) & 61,993 (39) & 51,616 (38) & 57,970 (38) \\ \hline
\end{tabular}
}
\end{table*}
The experiments were conducted using the CIFAR-100, providing a standard benchmark for assessing classification accuracy (100 classes, 50K training images, 10K test images) and three endoscopic databases, namely, the REAL-Colon \cite{realcolon2023}, Misawa et al.'s database \cite{misawa2021cad}, and the SUN dataset \cite{suncolon2020}, consisting of a total of 232 unique polyps, represented by 372,040 frames. This dataset is partitioned into six folds to facilitate k-fold cross-validation. The database's overview, detailed in Table \ref{tab:dataset_stats}, highlights the distribution of lesions in the dataset, revealing a significant imbalance among the various types of lesions; for more specifics, please see \cite{realcolon2023,misawa2021cad,suncolon2020}. For the sizing task, we used a 6-Fold cross-validation, allocating one fold each for testing and validation, and four folds for training in each cycle.\\

\noindent
\textbf{Exp. 1: Evaluating the ViT-Tiny and ResNet-18 on CIFAR-100}\\

% \begin{figure}[t]
%     \centering
%     \includegraphics[width=0.85\linewidth]{CIFAR_100.pdf}
% \caption{ViT/R18 Performance on CIFAR-100: in line with \cite{TinyViT2022,Deng2020R18Cifar}.}
%     \label{fig:cifar_results}
% \end{figure}
\begin{figure*}[t]
    \centering
    \begin{minipage}[c]{0.7\textwidth}
        \includegraphics[width=1.0\linewidth]{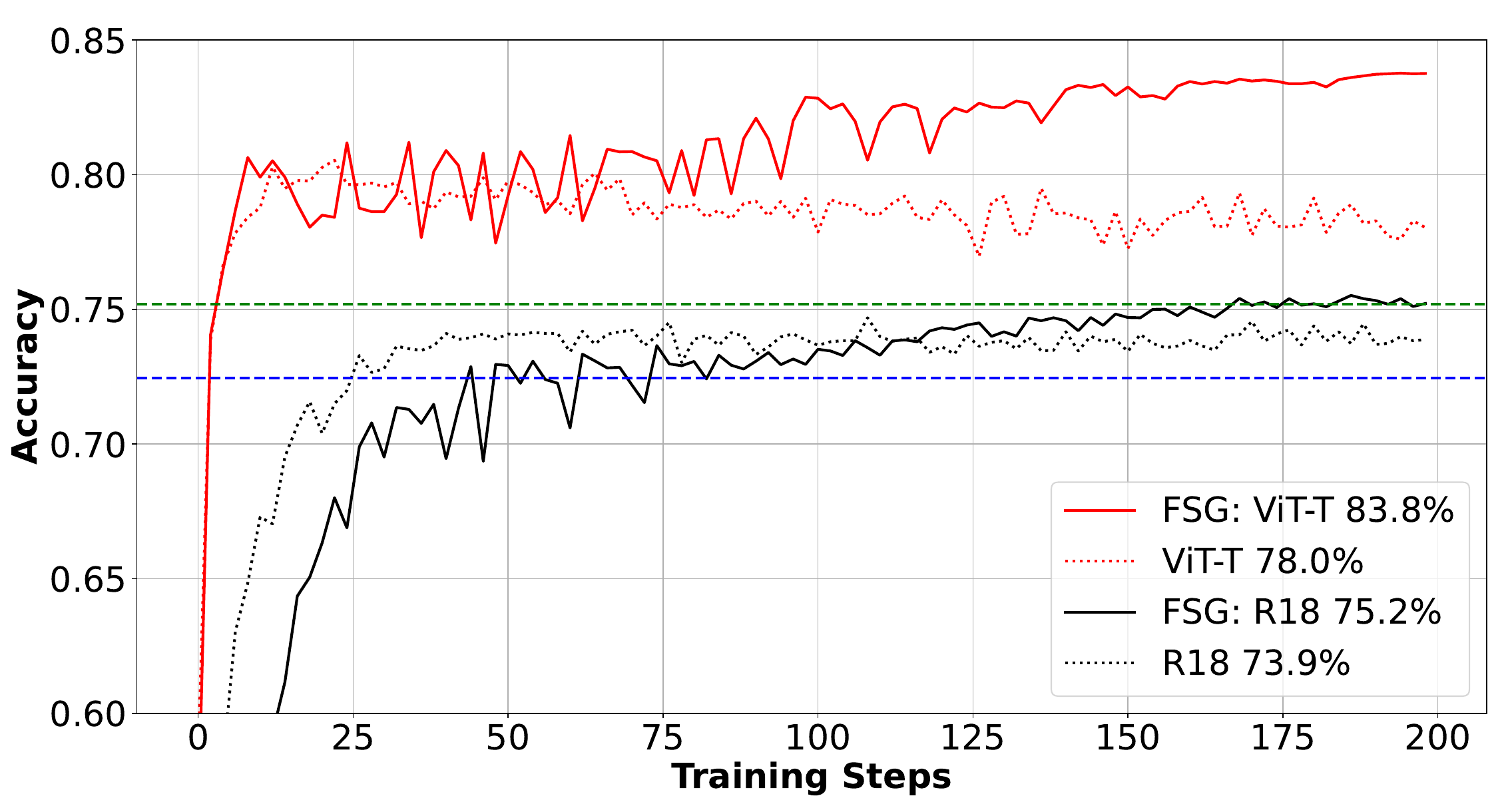} % Adjust height as needed
    \end{minipage}\hfill
    \begin{minipage}[c]{0.3\textwidth}
    \resizebox{0.75\textwidth}{!}{%
        \begin{tabular}{|c|c|}
            \hline
            \textbf{Methods} & \textbf{Acc.} \\
            \hline
            R18 in \cite{Deng2020R18Cifar} & 74.1\% \\
            R18 (re-train) & 73.9\% \\
            \textbf{R18+HAG} & \textbf{75.2\%} \\
            \hline
            ViT-T in \cite{TinyViT2022} & 75.2\% \\
            ViT-T (re-train) & 78.0\% \\
            \textbf{ViT-T+HAG} & \textbf{83.8\%} \\
            \hline
        \end{tabular}
        }
        % \caption{ViT/R18 result table.}
        % \label{tab:your_table_label}
    \end{minipage}
     \caption{ViT/R18 Performance on CIFAR-100 compared with SotA \cite{TinyViT2022,Deng2020R18Cifar}.}\label{fig:cifar_results}
\end{figure*}
\noindent
Our experimental analysis on the CIFAR-100 dataset highlights the significant impact of integrating HAGs with CNN and Transformer architectures, specifically ResNet-18 (R18) and Vision Transformer Tiny (ViT-T), on classification accuracy. Initially, the R18 model reached a 73.9\% accuracy, aligning with previous benchmarks~\cite{Deng2020R18Cifar}. Incorporation of HAG into R18 improved its accuracy to 75.2\%, indicating a 1.8\% improvement. For ViT-T, the initial accuracy stood at 78.0\%, comparable to standards set in~\cite{TinyViT2022}. However, applying HAG to ViT-T significantly increased its accuracy to 83.8\%, marking a substantial 5.8\% enhancement over the baseline. These improvements underscore HAG's capability to selectively emphasize influential features, thereby optimizing model performance across both architectures. This foundational assessment sets the stage for applying HAG in more specialized tasks like polyp size estimation, demonstrating its potential to refine accuracy in complex image classification challenges.\\

\noindent
\textbf{Exp. 2: Polyp-Size Estimation. Bias-Variance Tradeoff Analysis}\\
\begin{table*}[t]
\centering
\resizebox{\textwidth}{!}{%
\begin{tabular}{|l|c|c|c|c|c|c|c|c|c|}
\hline
\textbf{Method Name} & \multicolumn{2}{c|}{\textbf{Balanced Accuracy}} & \multicolumn{2}{c|}{\textbf{Avg. Sens-Spec}} & \multicolumn{2}{c|}{\textbf{F1 Score}} & \textbf{Global Average} & \textbf{Params} & \textbf{\#flops} \\
 & \textbf{Value} & \textbf{Variance} & \textbf{Value} & \textbf{Variance} & \textbf{Value} & \textbf{Variance} & \textbf{of Metrics} & & \\
\hline
R18 \cite{he2016deep,Deng2020R18Cifar}(RGB) & \textbf{53.4}\% & 1.4\% & 57.3\% & 1.0\% & 77.5\% & \textbf{1.8}\%  & 62.73\%& 11.2M & 5.3G\\
\textbf{HAG: R18 (RGB)} & 52.3\% & \textbf{0.7}\% & \textbf{57.9}\% & \textbf{0.5}\% & \textbf{78.7}\% & 2.0\%  & \textcolor{darkgreen}{\textbf{62.97}}\%& 11.3M  & 5.3G\\
\hline
R18 \cite{ranftl2020towards,MIDAS}(DPT) & \textbf{46.9}\% & 1.0\% & 46.8\% & 1.6\% & 66.2\% & 5.0\% & 53.30\%& 11.2M & 5.1G  \\
\textbf{HAG: R18 (DPT)} & 46.8\% & \textbf{0.4}\% & \textbf{50.1}\% & \textbf{0.3}\% & \textbf{76.5}\% & \textbf{2.1}\%  & \textcolor{darkgreen}{\textbf{57.80}}\%& 11.2M & 5.1G \\
\hline
R18 (LOC) & \textbf{54.3}\% & 1.7\% & \textbf{56.4}\% & 1.3\% & \textbf{78.1}\% & 2.3\%  & \textcolor{darkgreen}{\textbf{62.93}}\%& 11.2M &   5.3G\\
\textbf{HAG: R18 (LOC)} & 53.0\% & \textbf{1.2}\% & 55.4\% & \textbf{0.9}\% & 77.6\% & \textbf{1.7}\%  & 62.00\%& 11.3M  & 5.3G\\
\hline
MultiStream-R18 \cite{baltruvsaitis2018multimodal}[RGB+DPT] & 50.7\% & 1.3\% & 54.9\% & \textbf{0.8}\% & 77.0\% & \textbf{2.2}\%  & 60.87\% & 22.4M & 5.3G\\\
\textbf{HAG: MultiStream-R18 (RGB+DPT)} & \textbf{53.4}\% & 1.3\% & \textbf{57.1}\% & 0.9\% & \textbf{77.8}\% & 2.5\%  & \textcolor{darkgreen}{\textbf{62.77}}\% & 22.5M & 5.3G\\
\hline
MultiStream-R18 \cite{baltruvsaitis2018multimodal}[LOC+DPT] & 52.0\% & \textbf{0.7}\% & 53.1\% & \textbf{0.4}\% & 78.2\% & 1.7\%  & 61.10\% & 22.4M & 5.3G\\
\textbf{HAG: MultiStream-R18 (LOC+DPT)} & \textbf{53.5}\% & 0.9\% & \textbf{56.0}\% & 0.5\% & \textbf{79.5}\% & \textbf{1.4}\% & \textcolor{darkgreen}{\textbf{63.00}}\% & 22.5M &  5.3G\\
\hline
ViT-Tiny \cite{vit2021,TinyViT2022}(RGB) & 51.3\% & 1.2\% & 55.7\% & \textbf{0.6}\% & 75.6\% & 2.0\% & 60.87\% & \textcolor{darkgreen}{\textbf{5.6M}}  &  \textcolor{darkgreen}{\textbf{4.7G}}\\
\textbf{HAG: ViT-Tiny (RGB)} & \textbf{54.9}\% & \textbf{1.1}\% & \textbf{59.5}\% & 0.7\% & \textbf{79.1}\% & 2.0\%  & \textcolor{darkgreen}{\textbf{64.50}}\%& \textcolor{darkgreen}{\textbf{5.6M}} &  \textcolor{darkgreen}{\textbf{4.7G}}\\
\hline
\end{tabular}
}
\caption{Bias \& Variance: Diminutive (<=5mm), Small (5-10mm), Large (>=10mm)}
\label{tab:bias_variance_analysis}
\end{table*}

\noindent
In our experimental setup, models were trained on four folds, with one fold each for validation and testing, as per Table~\ref{tab:dataset_stats}. The optimal model checkpoint is chosen based on the lowest validation loss.

Table~\ref{tab:bias_variance_analysis} presents a detailed bias-variance tradeoff analysis for models enhanced with HAGs using RGB, LOC, and DPT inputs. The integration of HAG into the ViT model with RGB inputs increases Balanced Accuracy (BA) from 51.3\% to 54.9\%, improves Avg. Sensitivity-Specificity to 59.5\%, and raises the F1 Score to 79.1\%. This results in a global average metric improvement of +3.63\% relative to its non-HAG counterpart, achieving 64.5\%.
In the case of the R18 model equipped with DPT inputs, the integration of HAG results in an increase of Avg. Sensitivity-Specificity by +3.3\% and an F1 Score by +10.3\%, both compared to the model's performance without HAG. Applying HAG to the R18 model with LOC inputs results in a slight decrease in BA by approximately 1\%. The combination of RGB and polyp location masks, which utilize ground truth (GT) bounding boxes to create the location masks, represents an ideal input with optimal feature selection by design. In this scenario, HAG had limited scope for reweighting and selection as most relevant information was already incorporated. For multi-stream configurations combining LOC and DPT inputs, the enhancements include a BA increase to 53.5\% and the attainment of the highest F1 Score at 79.5\%.
These improvements are due to HAG and GR. HAG promotes sparse connectivity, reducing overfitting and improving generalization. GR optimizes HAG with dual forward passes, focusing on key features and eliminating redundancies when the main model parameters are frozen. This ensures the model focuses on relevant features, enhancing predictive accuracy and robustness across input modalities.\\

\noindent
\textbf{Exp. 3: Comprehensive Model Performance Analysis}\\
\begin{table*}[t]
\centering
\resizebox{1.0\textwidth}{!}{
\begin{tabular}{|l|ccc|c|ccc|c|c|}
\hline
& \multicolumn{4}{c|}{\textbf{Binary Classification}} & \multicolumn{4}{c|}{\textbf{Triclass Classification}} & \textbf{Overall}\\
\cline{2-9}
\textbf{Method Name} & \textbf{Bal. Acc.} & \textbf{F1 Score} & \textbf{Sens.-Spec.} & \textbf{Avg.} & \textbf{Bal. Acc.} & \textbf{F1 Score} & \textbf{Sens.-Spec.} & \textbf{Avg.} & \textbf{Score}\\
\hline
R18 \cite{he2016deep,Deng2020R18Cifar}(RGB) & \textbf{57.58\%} & \textbf{87.22\%} & \textbf{70.20\%} & 71.67\% & \textbf{47.62\%} & 74.84\% & \textbf{54.18\%} & 58.88\% & 65.2\%\\
\textbf{HAG: R18 (RGB)}      & 55.54\% & 86.46\% & \textbf{70.10\%} & 70.70\% & 45.57\% & \textbf{75.68\%} & \textbf{54.00\%} & 58.42\% & 64.6\%\\
\hline
R18 \cite{ranftl2020towards,MIDAS}(DPT) & \textbf{58.20\%} & 78.93\% & 56.51\% & 64.55\% & 44.19\% & 68.12\% &42.72\% & 51.68\% & 58.1\%\\
\textbf{HAG: R18 (DPT)}      & 54.75\% & \textbf{85.84\%} & \textbf{63.30\%} & 67.96\% & \textbf{45.0\%} & \textbf{73.58\%} & \textbf{49.00\%} & 55.86\% & 61.9\%\\
\hline
R18 (LOC)                    & 54.12\% & 85.59\% & 62.56\% & 67.42\% & 44.76\% & 74.99\% & 48.93\% & 56.23\% & 61.8\%\\
\textbf{HAG: R18 (LOC)}      & \textbf{55.70\%} & \textbf{86.24\%} & \textbf{64.80\%} & 68.91\% & \textbf{45.42\%} & 74.99\% & \textbf{50.11\%} & 56.84\% & 62.9\%\\
\hline
MultiStream-R18 \cite{baltruvsaitis2018multimodal}[RGB+DPT]    & \textbf{54.40\%} & \textbf{85.90\%} & \textbf{67.09\%} & 69.13\% & 44.36\% & 73.81\% & 51.34\% & 56.50\% & 62.8\%\\
\textbf{HAG: MultiStream-R18 [RGB+DPT]} & 53.63\% & 85.55\% & 65.49\% & 68.22\% & \textbf{45.81\%} & \textbf{74.36\%} & \textbf{52.01\%} & 57.39\% & 62.8\%\\
\hline
MultiStream-R18 \cite{baltruvsaitis2018multimodal}[LOC+DPT]    & 55.92\% & 86.16\% & 63.36\% & 68.48\% & 47.00\% & 75.68\% & 50.25\% & 57.64\% & 63.1\%\\
*\textbf{HAG: MultiStream-R18 [LOC+DPT]} & \textbf{59.96\%} & \textbf{87.84\%} & \textbf{69.23\%} & \textcolor{darkgreen}{\textbf{72.34\%}} & \textbf{48.84\%} & \textbf{77.11\%} & \textbf{53.56\%} & \textcolor{darkgreen}{\textbf{59.84\%}} & \textcolor{darkgreen}{\textbf{66.1\%*}}\\
\hline
ViT-Tiny \cite{vit2021,TinyViT2022}(RGB) & 54.63\% & 86.04\% & 68.35\% & 69.67\% & 42.78\% & 72.80\% & 50.82\% & 55.47\% & 62.5\%\\
*\textbf{HAG: ViT-Tiny (RGB)}    & \textbf{55.86\%} & \textbf{86.56\%} & \textbf{69.33\%} & 70.58\% & \textbf{48.93\%} & \textbf{76.47\%} & \textbf{55.62\%} & \textcolor{darkgreen}{\textbf{60.34\%}} & \textcolor{darkgreen}{\textbf{65.5\%*}}\\
\hline
\end{tabular}
}
\caption{\textbf{Performance Summary}: \textit{Binary}~\cite{itoh2021binary} (Polyps < 10mm vs. >= 10mm) \textbf{vs} \textit{Triclass Classification} (Diminutive: <= 5mm, Small: 5-10mm, Large: >= 10mm)}
\label{tab:combined_performance_summary}
\end{table*}

\noindent
In our analysis, models enhanced with HAGs were evaluated for generalizability across \textit{binary} and \textit{triclass} classifications, as detailed in Table \ref{tab:combined_performance_summary}. In this experiment, we consolidated all inferences across different folds for each model to provide a comprehensive overview of their performance. 

\textbf{In binary classification tasks}, the HAG MultiStream-R18 (LOC+DPT) model achieved a Balanced Accuracy (BA) of 59.96\%, an F1 Score of 87.84\%, and a Sensitivity-Specificity of 69.23\%, with an overall average performance of 72.34\%. This model showcases the efficacy of HAG in improving precision and predictive accuracy, setting a high benchmark in the binary classification domain.

\textbf{In the context of triclass classification}, the task's complexity significantly escalates. Nonetheless, the HAG-enhanced ViT-T (RGB) model showcases notable performance, achieving a Balanced Accuracy (BA) of 48.93\%, an F1 Score of 76.47\%, and a Sensitivity-Specificity of 55.62\%, culminating in an average of 60.34\%. These metrics not only underscore the model's robustness but also its adaptability to more complex classification scenarios, despite having only 5M parameters. This is considerably less — four times fewer than the multistream networks and half that of the R18.

The performance of the HAG-enhanced models, particularly MultiStream-R18 (LOC+DPT) in binary classification and ViT-T (RGB) in triclass classification, underscores the efficacy of HAG in model optimization across different classification tasks. Comparing 12 methods, the highest performances were from \textbf{HAG MultiStream-R18 (LOC+DPT)} at \textbf{66.1\%} and \textbf{HAG ViT-T (RGB)} at \textbf{65.5\%}. ViT models without LOC maps or DPT are preferred, due to their lower probability of error propagation from detection and depth estimation frameworks. Moreover, ViT-Tiny has four times fewer parameters and lower FLOPs (5.6M vs. 22.5M and 4.7G vs. 5.3G) compared to MultiStream-R18 (LOC+DPT). Therefore, the most promising solution is \textit{HAG ViT-Tiny} as shown in this paper; Vision Transformer models improve significantly in both CIFAR-100 (classification in natural imaging) and polyp size estimation (regression in medical imaging).

\section{Conclusions}

This study advances deep learning for polyp size assessment by innovatively integrating Feature Selection-Attention Gates (HAG) with Gradient Routing (GR) across CNN and ViT architectures. For polyp sizing, the HAG-enhanced MultiStream-R18 (LOC+DPT) model excels in binary classification, achieving an F1 Score of 87.8\% and an average performance of 72.3\%. In triclass classification, the ViT-T model attains an F1 Score of 76.5\% and an average of 60.3\%, highlighting its efficiency and adaptability. Furthermore, the HAG-enhanced ViT achieves 83.8\% accuracy in CIFAR-100, demonstrating its versatility for various imaging tasks. ViT models without LOC maps or DPT are preferred due to their lower probability of error propagation in detection and depth estimation frameworks. Moreover, ViT Tiny, with 5.6M params and 4.7G flops, has the lowest parameter count. Integrating HAG enhances ViT, achieving top performance in regression and classification. Advanced Vision Transformers like Swin, DeiT, and PVT show significant potential for future research. This analysis emphasizes the pivotal role of HAG-GR in improving polyp size estimation, suggesting beneficial effects on clinical outcomes. We aim to expand the application of these techniques to a broader range of medical imaging challenges, improving diagnostic accuracy with minimal computational overhead. To facilitate future research, the dataset splits and codebase for CNNs, multistream CNNs, ViT, and HAG-enhanced models is available at \href{http://github.com/cosmoimd/feature-selection-gates}{github.com/cosmoimd/feature-selection-gates}.\\\\

\noindent
\textbf{Disclosure of Interests.} All the authors are affiliated with Cosmo Intelligent Medical Devices, the developer of the GI Genius medical device.

%
% ---- Bibliography ----

% \end{thebibliography}

\section{Appendix}

\section{HAG Weight Distributions in the ViT Model}

Our research highlights the adaptability of Vision Transformer (ViT) architectures enhanced with Hard-Attention Gates (HAG) for tasks like CIFAR-100 classification and polyp sizing. CIFAR-100 is a classification task on natural images, while polyp sizing is a regression task on medical images. HAG, which are online feature re-weighting gates, act as a sort of hard-attention mechanism by scaling irrelevant dimensions and embeddings. This attenuates and reduces the model parameters without removing them. HAG dynamically adjusts feature importance based on task requirements, showcasing effective feature selection.

In CIFAR-100 classification, HAG prioritizes a range of features from basic (edges, lines) in early layers to complex (patterns, objects) in deeper layers. Weight distributions shift towards higher significance (initially $\mu=0.84$, $\sigma=0.13$, final layers $\mu=0.92$, $\sigma=0.07$), reflecting the need for comprehensive feature integration to classify diverse objects accurately. For polyp sizing, a regression task, HAG weights remain uniformly distributed ($\mu \approx 0.5$, $\sigma$ from 0.026 to 0.029), similar to the model's initial state. This uniformity suggests equitable feature consideration, essential for size and shape differentiation, akin to regularization methods like L1 and L2 that prevent overfitting.

The different HAG weight behaviors in these tasks are due to the nature of the tasks themselves. In classification tasks like CIFAR-100, the need to distinguish between many classes requires HAG to emphasize a wide range of features, aligning with hierarchical feature learning where early layers capture generic features, and deeper layers capture task-specific details. In regression tasks like polyp sizing, the goal is to predict a continuous value based on subtle differences in features such as size and shape. The uniform HAG weights ensure a balanced consideration of all features, minimizing regression error without overemphasis on specific features, similar to the effect of regularization techniques that prevent overfitting. In summary, HAG's adaptive feature moderation optimizes performance by emphasizing critical features in classification tasks and maintaining balanced feature integration in regression tasks, aligning with task-specific requirements. Additionally, optimizing HAG parameters separately from the main model allows tailored learning rates and gradient clippings, thus enhancing HAG networks' training efficiency.
\begin{figure}[t]
    \centering
    \includegraphics[width=0.8\linewidth]{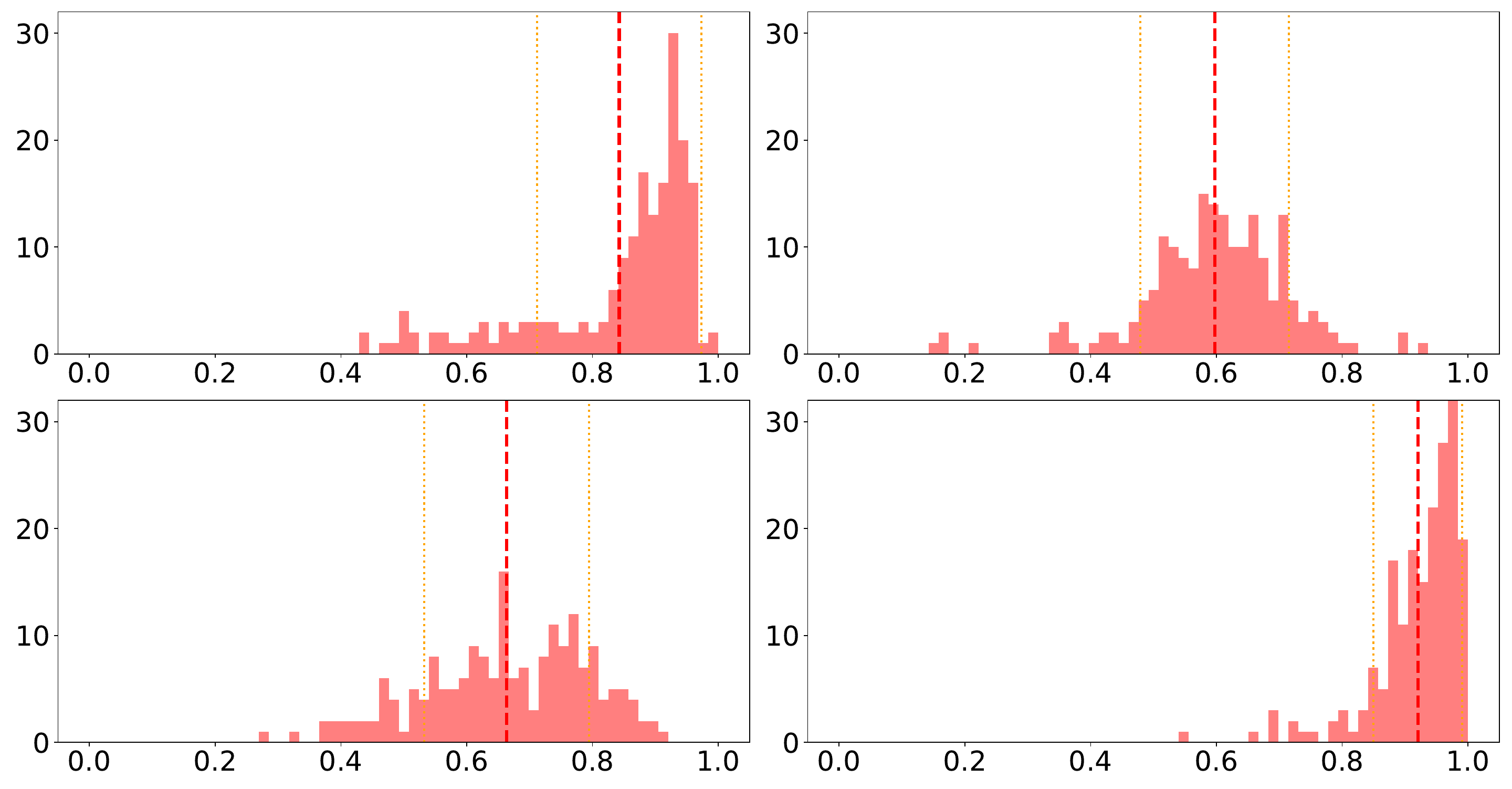}
    \put(-204,100){
    \tiny\textbf{ViT $Block_{0}$}}
    \put(-204,90){
    \tiny After-SA}
    \put(-100,100){
    \tiny\textbf{ViT $Block_{0}$}}
    \put(-100,90){
    \tiny After-MLP}
    \put(-204,48){
    \tiny\textbf{ViT $Block_{11}$}}
    \put(-204,38){
    \tiny After-SA}
    \put(-100,48){
    \tiny\textbf{ViT $Block_{11}$}}
    \put(-100,38){
    \tiny After-MLP}
    \caption{CIFAR-100: FSG-GR weight distributions in the ViT.}
    \label{fig:FSG_GR_ViT_CIFAR}
\end{figure}
\begin{figure}[t]
    \centering
    \includegraphics[width=0.8\linewidth]{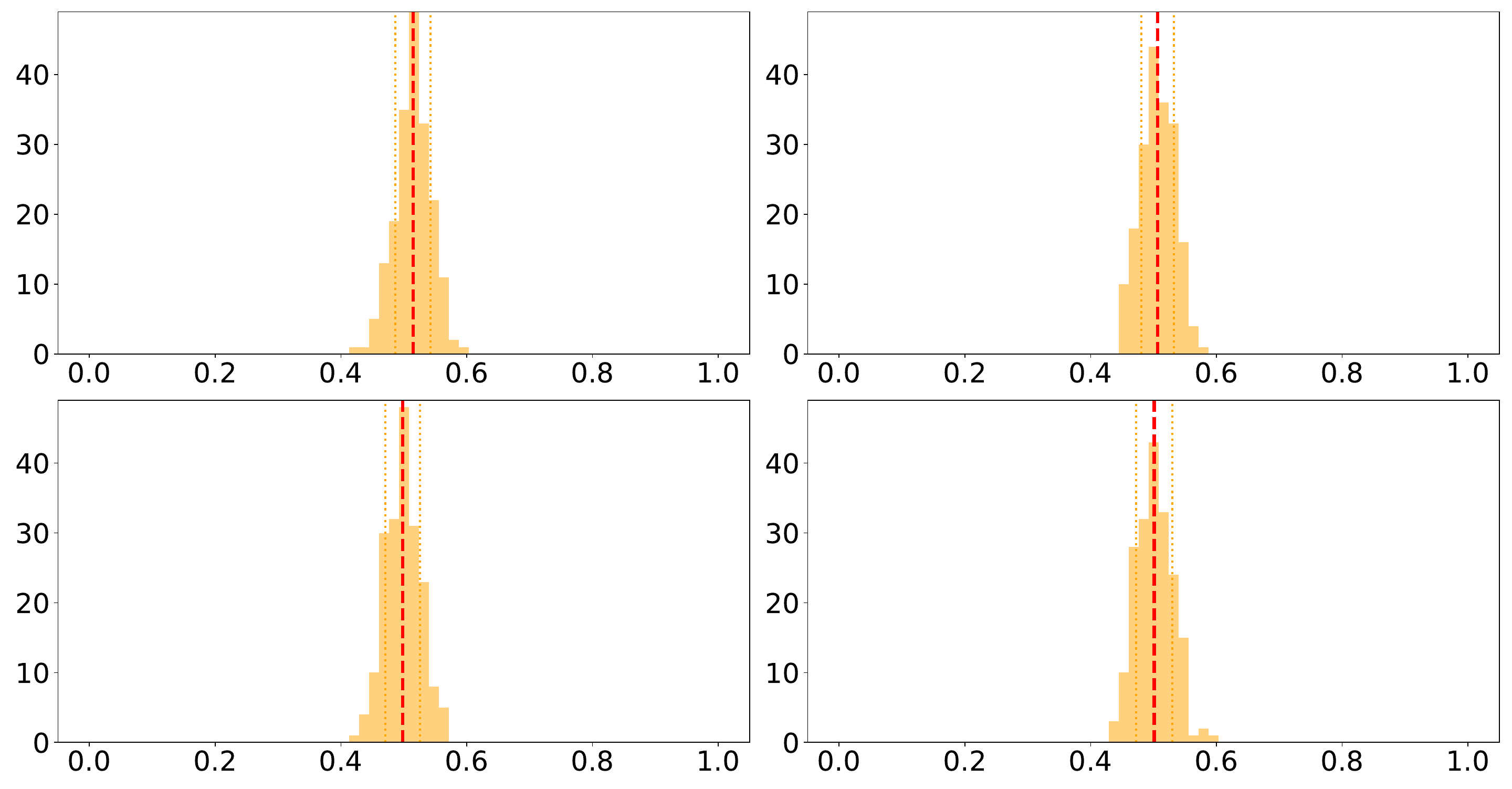}
    \put(-204,100){
    \tiny\textbf{ViT $Block_{0}$}}
    \put(-204,90){
    \tiny After-SA}
    \put(-100,100){
    \tiny\textbf{ViT $Block_{0}$}}
    \put(-100,90){
    \tiny After-MLP}
    \put(-204,48){
    \tiny\textbf{ViT $Block_{11}$}}
    \put(-204,38){
    \tiny After-SA}
    \put(-100,48){
    \tiny\textbf{ViT $Block_{11}$}}
    \put(-100,38){
    \tiny After-MLP}
    \caption{Polyp Sizing: FSG-GR weight distributions in the ViT.}
    \label{fig:FSG_GR_ViT_PSIZING}
\end{figure}
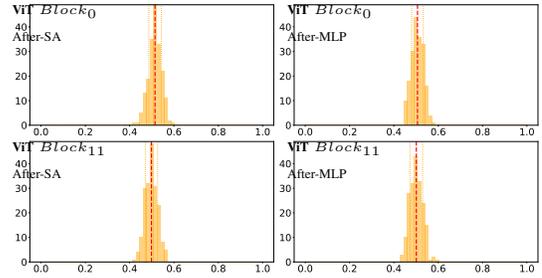
\section{Distribution of Frames and Polyps Across the 6 Folds}

In this appendix, we provide a detailed visualization and analysis of the distribution of frames and polyps across the six folds used in our study. This comprehensive breakdown is crucial for understanding the data balance and ensuring the reliability of our experimental results. 

\begin{figure*}[!]
    \centering
    \includegraphics[width=0.85\textwidth]{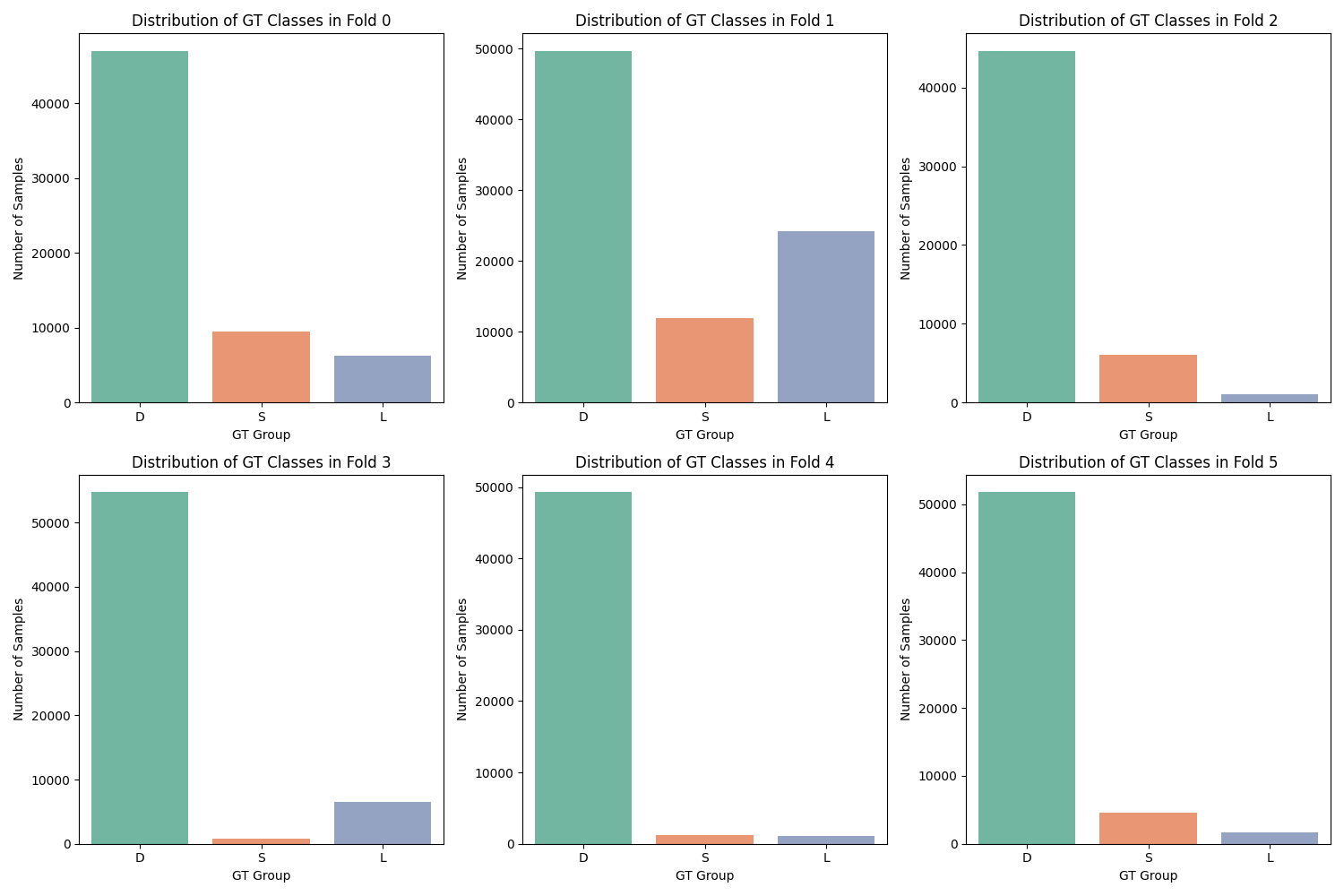}
    \caption{Distribution of GT Classes in each Fold.}
    \label{fig:gt_group_distribution_per_fold}
\end{figure*}

\begin{figure*}[!]
    \centering
    \includegraphics[width=0.85\textwidth]{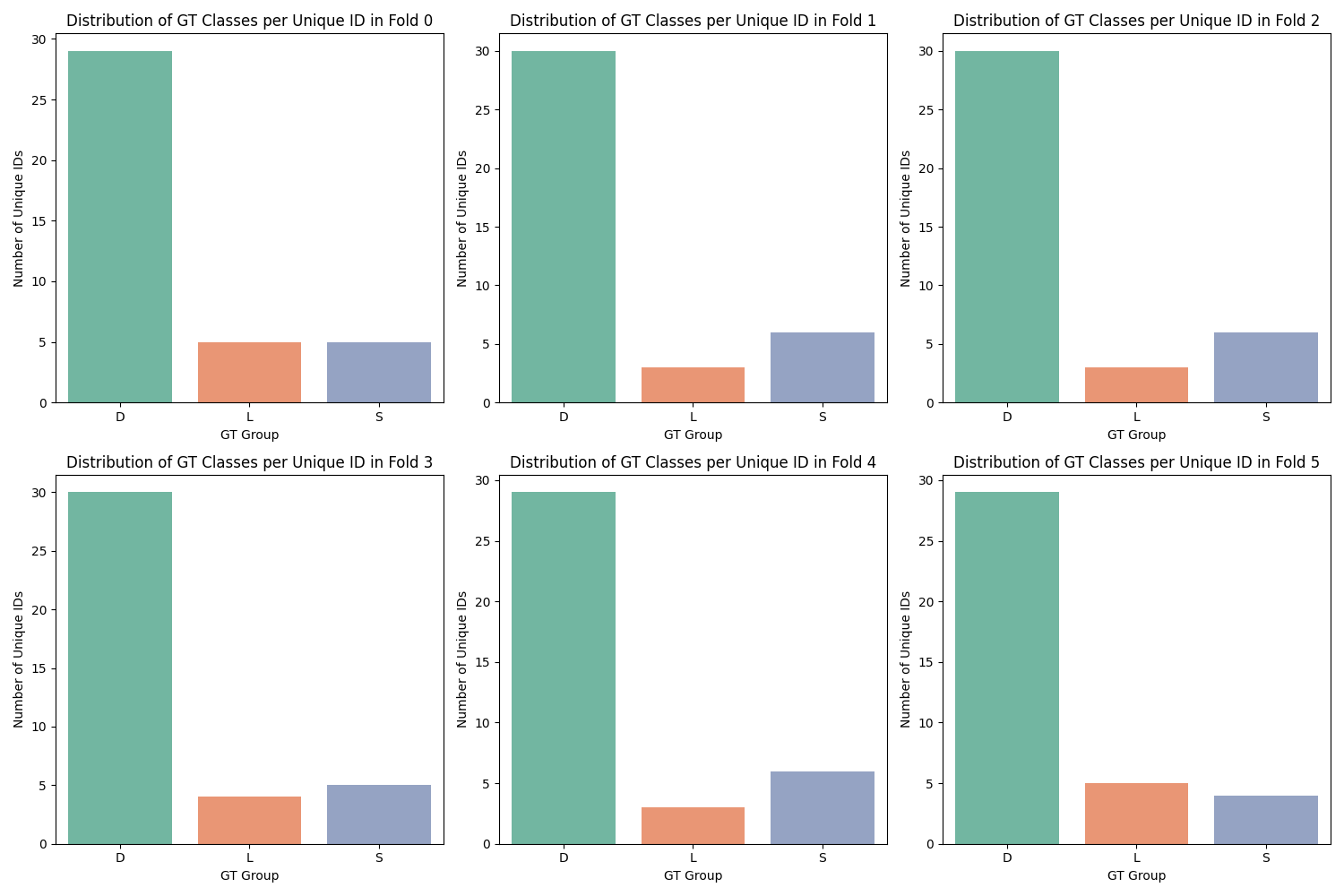}
    \caption{Distribution of GT Classes per Unique ID in each Fold.}
    \label{fig:gt_group_distribution_per_unique_id_per_fold}
\end{figure*}

\begin{figure*}[!]
    \centering
    \includegraphics[width=0.85\textwidth]{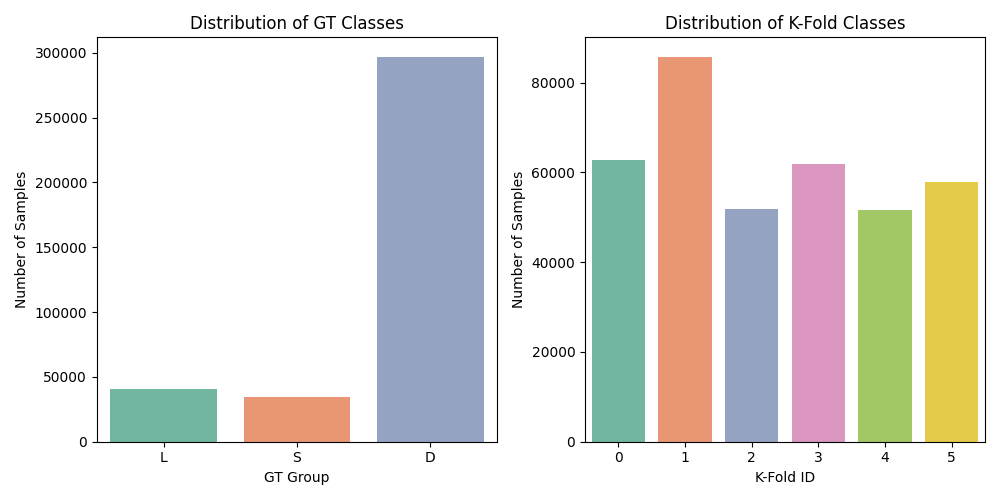}
    \caption{Overall Distribution of GT Classes and K-Fold Classes.}
    \label{fig:kfold_distribution}
\end{figure*}

Figures \ref{fig:gt_group_distribution_per_fold}, \ref{fig:gt_group_distribution_per_unique_id_per_fold}, and \ref{fig:kfold_distribution} illustrate various aspects of the dataset distribution, while Tables \ref{tab:dataset_stats} and \ref{tab:polyps_distribution} provide a numerical summary of the data.

Figure \ref{fig:gt_group_distribution_per_fold} shows the distribution of ground truth (GT) classes in each fold. This figure categorizes the data into three groups: Diminutive (D), Small (S), and Large (L) polyps. Each subplot represents one fold, showing the number of samples in each GT group. This visualization helps identify the balance or imbalance in the dataset, which is critical for training robust models. From the figure, we observe that the majority of the samples belong to the Diminutive category across all folds, with significantly fewer samples in the Small and Large categories. This imbalance needs to be considered when designing the models to ensure they can generalize well across all classes.

\begin{table*}[!]
\centering
\caption{Distribution of Frames (\#Polyps) in Endoscopic Dataset Folds from REAL-Colon \cite{realcolon2023}, Misawa et al.'s Database \cite{misawa2021cad}, and the SUN Dataset \cite{suncolon2020}}
\label{tab:dataset_stats}
\resizebox{0.85\textwidth}{!}{%
\begin{tabular}{|c|c|c|c|c|c|c|}
\hline
\textbf{Categories} & \textbf{Fold 1} & \textbf{Fold 2} & \textbf{Fold 3} & \textbf{Fold 4} & \textbf{Fold 5} & \textbf{Fold 6} \\ \hline
\textit{Diminutive} & 46,982 (29) & 49,648 (30) & 44,640 (30) & 54,707 (30) & 49,256 (29) & 51,784 (29) \\ \hline
\textit{Small} & 9,535 (5) & 11,985 (6) & 6,098 (6) & 827 (5) & 1,255 (6) & 4,538 (4) \\ \hline
\textit{Large} & 6,298 (5) & 24,189 (3) & 1,086 (3) & 6,459 (4) & 1,105 (3) & 1,648 (5) \\ \hline
\textbf{Total} & 62,815 (39) & 85,822 (39) & 51,824 (39) & 61,993 (39) & 51,616 (38) & 57,970 (38) \\ \hline
\end{tabular}
}
\end{table*}
\begin{table*}[!]
\centering
\caption{Distribution of Frames and Polyps in the Dataset}
\label{tab:polyps_distribution}
\resizebox{0.6\textwidth}{!}{%
\begin{tabular}{|c|c|c|}
\hline
\textbf{GT Group} & \textbf{Number of Samples} & \textbf{Number of Unique IDs} \\ \hline
\textit{Diminutive (D)} & 300,000+ & 30 \\ \hline
\textit{Small (S)} & 50,000+ & 10 \\ \hline
\textit{Large (L)} & 30,000+ & 5 \\ \hline
\end{tabular}
}
\end{table*}
Figure \ref{fig:gt_group_distribution_per_unique_id_per_fold} depicts the distribution of GT classes per unique ID in each fold. Similar to the previous figure, it classifies the data into Diminutive, Small, and Large groups. However, this figure focuses on the number of unique IDs rather than the total number of samples. Each subplot corresponds to one fold, showing the number of unique polyps in each GT group. This visualization highlights how the data is distributed in terms of individual polyps, ensuring that the folds are not only balanced in terms of sample count but also in terms of unique polyps. It helps verify that no single fold is disproportionately represented by certain polyps, which could skew the model's learning process.

Figure \ref{fig:kfold_distribution} presents the overall distribution of GT classes and the distribution of K-Fold classes. The left subplot shows the cumulative distribution of samples across all folds, reinforcing the observation that the dataset is predominantly composed of Diminutive polyps. The right subplot displays the number of samples in each fold, providing an overview of how the data is partitioned. This figure ensures that each fold has a comparable number of samples, which is essential for fair cross-validation.

Table \ref{tab:dataset_stats} summarizes the distribution of frames and polyps across the six folds from the REAL-Colon, Misawa et al.'s database, and the SUN dataset. It provides a numerical breakdown of the number of frames and polyps in each GT category (Diminutive, Small, Large) for each fold. This table complements the visualizations by providing exact counts, aiding in a precise understanding of the dataset composition. Table \ref{tab:polyps_distribution} provides an additional layer of detail by summarizing the overall number of samples and unique IDs in each GT group across the entire dataset. This table highlights the significant imbalance in the dataset, with a substantial majority of samples classified as Diminutive. This information is crucial for interpreting the performance metrics of the models and for considering strategies to address class imbalance, such as data augmentation or weighted loss functions. In conclusion, the figures and tables presented in this appendix offer a comprehensive overview of the dataset distribution. They highlight key aspects such as class imbalance and the partitioning of data across folds, which are vital for understanding the challenges and ensuring the robustness of the experimental evaluations in our study.

\section{Experimental Settings}
\begin{table*}[!]
\centering
\caption{Experimental Settings for Models with and without HAG}
\label{tab:experimental_settings}
\resizebox{0.85\textwidth}{!}{%
\begin{tabular}{|c|c|c|c|}
\hline
\multirow{2}{*}{\textbf{Category}} & \multirow{2}{*}{\textbf{Parameter}} & \multicolumn{2}{c|}{\textbf{Experimental Settings}} \\ \cline{3-4} 
                                   &                                     & \textbf{With HAG}     & \textbf{Without HAG}    \\ \hline
\multirow{3}{*}{Preprocessing}     & Image Resizing                      & \multicolumn{2}{c|}{$384 \times 384$ pixels} \\ \cline{2-4} 
                                   & Normalization Means                 & \multicolumn{2}{c|}{0.32239652, 0.22631808, 0.17500061} \\ \cline{2-4} 
                                   & Normalization STDs                  & \multicolumn{2}{c|}{0.31781745, 0.2405859, 0.19327126} \\ \hline
Augmentation                       & Techniques                          & \multicolumn{2}{c|}{Random Rotations, Color Jittering, Gaussian Noise} \\ \hline
\multirow{2}{*}{Target Norm.}      & Range                               & \multicolumn{2}{c|}{$[-1, +1]$} \\ \cline{2-4} 
                                   & Min-Max Values                      & \multicolumn{2}{c|}{0.5 mm, 20.0 mm} \\ \hline
Data Balancing                     & Frame Selection                     & \multicolumn{2}{c|}{40-128 Frames per Video} \\ \hline
\multirow{3}{*}{Loss Function}     & Type                                & Weighted Huber       & Weighted Huber       \\ \cline{2-4} 
                                   & Parameters Model                    & Weight [3,5], Threshold [5,10]   & Weight [3,5], Threshold [5,10]  \\ 
                                   & Parameters HAG                      & Weight [5,10], Threshold [5,10]  &  N.A. \\ \hline
\multirow{4}{*}{Training}          & Optimizer                           & \multicolumn{2}{c|}{Adam} \\ \cline{2-4} 
                                   & Learning Rate                       &  \(1 \times 10^{-2}\)-\(1 \times 10^{-5}\)   & \(1 \times 10^{-3}\)\\ \cline{2-4} 
                                   & Weight Decay                        &  \(1 \times 10^{-2}\)-\(1 \times 10^{-8}\) & \(1 \times 10^{-5}\) \\ \cline{2-4}
                                   & Gradient Clipping                   & 64.0-128.0     & 5.0-8.0   \\ \hline
\end{tabular}
}
\end{table*}
In this section, we outline the experimental settings used for training and evaluating the models with and without the Hard-Attention Gates (HAG). These settings ensure a fair comparison between the different model configurations.

Images were resized to \(384 \times 384\) pixels and normalized using dataset-specific mean and standard deviation values computed on the training data. The normalization means were set to (0.32239652, 0.22631808, 0.17500061) and the standard deviations to (0.31781745, 0.2405859, 0.19327126). This preprocessing ensures consistency across different datasets and corrects for color variations. Circular cropping was employed to isolate the central part of each image, focusing the analysis on relevant regions and minimizing distractions from the periphery.

Standard data augmentation techniques were applied, including random rotations, color jittering, and Gaussian noise. These augmentations help to improve the robustness and generalization ability of the models by introducing variability in the training data.

The target polyp sizes were normalized to a \([-1, +1]\) range to stabilize the regression training process. The minimum and maximum values used for this normalization were 0.5 mm and 20.0 mm, respectively. Data balancing was performed by selecting 40-128 frames per video to ensure a representative distribution of samples across different videos.

The loss function used was a domain-specific weighted Huber Loss, designed to address the imbalanced distribution of polyp sizes within the dataset. The weighting parameters for the loss function were set as follows: \(\alpha_1 = 2\) for polyps between 5 mm and 10 mm, and \(\alpha_2 = 3\) for polyps larger than 10 mm. The loss function was applied to both the models with and without HAG.

For training, the Adam optimizer was used, with learning rates ranging from \(1 \times 10^{-2}\) to \(1 \times 10^{-5}\) for models with HAG, and a fixed learning rate of \(1 \times 10^{-3}\) for models without HAG. Weight decay was set between \(1 \times 10^{-2}\) and \(1 \times 10^{-8}\) for models with HAG, and \(1 \times 10^{-5}\) for models without HAG. Gradient clipping thresholds were set to 64.0-128.0 for models with HAG and 5.0-8.0 for models without HAG to prevent gradient explosion and ensure stable training.

These experimental settings were chosen to provide a robust framework for evaluating the impact of Hard-Attention Gates on model performance, ensuring consistency and fairness in the comparison between models with and without HAG.\\\\\\

\noindent
\textbf{Code and Datasets.} To facilitate further research, we are releasing our codebase, which includes implementations for CNNs, multistream CNNs, ViT, and HAG-augmented variants. This resource aims to standardize the use of endoscopic datasets, providing public training-validation-testing splits for reliable and comparable research in gastroenterological polyp size estimation. The codebase is available at \href{http://github.com/cosmoimd/feature-selection-gates}{github.com/cosmoimd/feature-selection-gates}.\\\\

\noindent
\textbf{Disclosure of Interests.} All the authors are affiliated with Cosmo Intelligent Medical Devices, the developer of the GI Genius medical device.\\\\

\end{document}